# Nitrogen incorporated Zinc oxide thin film for efficient ethanol detection

P. K. Shihabudeen[1], Mina Yaghoobi Notash[2,4], Jaber Jahanbin Sardroodi[3,4] and Ayan Roy Chaudhuri[1]∗

1. Materials Science Centre, Indian Institute of Technology Kharagpur, 721302 Kharagpur, West Bengal, India

2. Department of Physics, Faculty of Basic Sciences, Azarbaijan Shahid Madani University, Tabriz, Iran

3. Department of chemistry, Faculty of Basic Sciences, Azarbaijan Shahid Madani University, Tabriz, Iran

4. Molecular Simulation Laboratory (MSL), Azarbaijan Shahid Madani University, Tabriz, Iran

**Abstract**

Zinc oxide which is a n-type semiconducting metal oxide (SMO) has been a promising material for detecting ethanol vapor. However, pure ZnO based ethanol sensors often suffer from high working temperature, cross sensitivity towards methanol and poor stability against humidity. Doping ZnO with various metal ions has been widely explored as a proficient approach to improve its ethanol sensing properties, while anionic dopants have been rarely considered. Here in we demonstrate the effect of nitrogen doping on the ethanol sensing characteristics of ZnO thin films. Nitrogen doped ZnO (N-ZnO) thin films have been synthesized following sol-gel technique with

∗ Corresponding author: Tel: +91-3222-283978

Electronic mail: ayan@matsc.iitkgp.ac.in (Ayan Roy Chaudhuri)

urea as nitrogen precursor. Ethanol sensing characteristics of the N-ZnO thin film has been compared with pure ZnO sensor over a wide range of temperature and relative humidity conditions. The N-ZnO sensor exhibits significantly large ethanol sensing response at a lower operating temperature (~99 % at 225 °C vs ~81 % at 250 °C for ZnO), faster response time (12 s vs 33 s for ZnO), long term stability, improved resilience against humidity and selectivity towards ethanol over methanol and acetone. The experimental observations have been supplemented by estimating the adsorption energies of ethanol on ZnO and N-ZnO surface using density functional theory (DFT) calculations. We discuss that the microscopic origin of improved ethanol sensing of N-ZnO is related to the facile adsorption of ethanol molecules on the oxide surface which is promoted by modification of electronic properties of ZnO by the nitrogen dopant atoms.

## 1. Introduction

Ethanol has been an important volatile organic compound (VOC) that has widespread use in various domestic and industrial applications, such as precursor/solvents in chemical industries, fuel or fuel additives, manufacture of drugs, plastics, polishes, cosmetics etc., as well as the prime component of alcoholic beverages. Utilisation of ethanol also poses several safety concerns. For example, augmented usage of ethanol in industry increases the risk of explosion hazards [1] and ground-water pollution [2]. Further, driving motor vehicles under the influence of alcohol has remained a leading cause of road accidents around the world. Additionally, ethanol content in breath is also linked to pathological conditions related to liver [3,4]. It is thus pertinent to develop highly sensitive, selective and fast ethanol sensors for maintaining industrial safety as well as for non-invasive monitoring of human health conditions. Developing ethanol sensors based on chemiresistive semiconducting metal oxide (SMO) thin films have gained significant research attention in the recent past. SMO thin films offer several advantages, such as high sensitivity, low fabrication costs, and compatibility with standard semiconductor technology [5,6]. Various SMOs, both n-type (e.g. ZnO, $In_2O_3$ $SnO_2$, $WO_3$) and p-type (NiO, CuO, $CoO_3$, and $Fe_2O_3$) have been investigated for ethanol sensing applications [7]. Among different SMOs, ZnO which is an n-type oxide with a bandgap of ~3.2 eV, offers several advantages such as ease of synthesis, cost-effectiveness, excellent chemical and thermal stability[8,9]. Despite all these advantages, ZnO based gas sensors suffer from several drawbacks such as cross-sensitivity with other VOCs, high operating temperature, and poor stability against humidity [10–12].

In order to circumvent the issues related to the ethanol sensing performance of ZnO thin films, various approaches have been employed, such as control of microstructure, fabrication of nano-composite thin films with other metal oxides, loading the oxide layer with nobles metals, incorporation of dopants etc [13–18]. Doping is a simple and effective method to improve the sensitivity and selectivity of SMO based gas sensors.Incorporation of suitable dopants modifies the electronic structure, disrupt chemical bonding, increases oxygen vacancy concentration and induces lattice stress in the SMO crystals which often leads to preferential adsorption of selective gas molecules [19,20]. In case of ZnO, incorporation of various metallic dopants such as In, Sn, Ce, Ni etc. have been found to improve its gas sensing characteristics [18–21]. However, metal dopants often induce thermal instability and carrier recombination sites which greatly limits performance of the oxide layer [22]. In this context, incorporation of non-metal dopants into ZnO can be a suitable alternative. Qin *et al.* and J. J. Macías-Sánchez *et al.* reported enhanced photocatalytic properties in nitrogen doped ZnO [23,24]. However, anion site doping is rarely explored for gas sensing properties of ZnO. Recently, Wen *et al.* reported VOC sensing properties of nitrogen doped ZnO powders where enhanced ethanol sensing characteristics have been attributed to enhanced oxygen vacancy concentration due to nitrogen doping [25]. However, the microscopic mechanism of improved ethanol sensing in nitrogen doped ZnO remains to be clarified. Further, to the best of our knowledge impact of nitrogen incorporation on the ethanol sensing properties of zinc oxide thin films have not been reported so far. Here in with we report the influence of substitutionally doped nitrogen on ethanol sensing of ZnO thin film in comparison with undoped ZnO thin film. We demonstrate that nitrogen doping in ZnO enhances its sensitivity and selectivity towards ethanol vapour. Additionally, the response time is reduced and stability against humidity is improved. Using DFT calculations we demonstrate that the microscopic origin of improved ethanol sensing of N-ZnO is related to the facile adsorption of ethanol molecules on the oxide surface which is promoted by nitrogen dopant atoms.

## 2. Experimental

2.1 Synthesis and materials

Zinc acetate dihydrate (Merk, >99 %) and urea (Alpha-aeser, >99 %) were used as a precursor for zinc and nitrogen, respectively, for the synthesis of nitrogen doped ZnO thin film. Firstly, the sol is synthesized, for which a solution of Zinc acetate (1.0975g) was prepared in 10 ml of ethanol (99.9 %). To which 3 ml of stabilizing agent, Ethanolamine, was added dropwise while stirring at 60 °C. Then, under similar stirring conditions, the desired amount of urea,1.2 g, was dissolved in a separate beaker. Later these solutions were blended in a separate beaker, and it was stirred at 60 °C for 4 hours till a stable reddish-orange solution was obtained [26]. For the synthesis of pure zinc oxide sol, a similar process was followed, without urea. The thin film was prepared using the spin coating technique. A thermally oxidized silicon wafer, cleaned to remove organic residue, is used as the substrate. After the substrate was washed and dried, as prepared sol of N-ZnO was spin-coated on it at 3000 rpm. After coating, the solvents were removed by drying them on a hot plate at 125 °C. The N-ZnO film with the desired thickness was prepared by spin coating the sol four times. The prepared film was annealed for one hour at 500 °C to obtain a crystalline N-ZnO thin film. Similarly, a reference sample of pure zinc oxide thin film was synthesized with the procedure mentioned earlier. The schematic representation for synthesis of N-ZnO thin film is portrayed in scheme S1.

2.2 Material characterisation

The crystal structures of the synthesized materials were determined by Panalytical Empyrion X-ray diffraction (XRD) Machine with a Cu K-α radiation source (1.54 Å). The microstructure and cross-section of the obtained thin films were analysed using a field emission scanning electron microscope (FESEM Zeiss Gemini, Germany). The surface chemical composition was studied by X-ray photoelectron spectroscopy (model PHI 5000 Versa Probe II, INC, Japan) with A1 K-α x-ray source (1486.6 eV). The UV–vis diffuse reflectance spectroscopy (UV–vis DRS) study were carried out with a Cary 5000 UV–vis spectrophotometer (Agilent Tech.)

2.3 Sensor measurements

In this study, we explored the suitability of N-ZnO thin film as an ethanol sensor. Interdigitated gold electrodes (with electrode separation of 1.5 mm and width of 2 mm) were sputtered on thin films for gas sensing studies. The VOC sensing behaviour of the thin film has been evaluated using a quasi-static gas sensing system. The sensing element is connected in series with a load resistance ($R_L$); both have similar resistance. A constant ~5 V DC is maintained through the sensor. A microcontroller-based (Atmel ATMEGA 32) data acquisition arrangement is used to measure the net output voltage across $R_L$. For further analyses and data storage, the whole data acquisition unit is connected to a desktop computer via the RS232 interface. The detailed measurement process could be found in our previous work[26]. The sensor response (S) of the thin films were calculated using the following relation

$$S(\%) = \frac{(R_a - R_g)}{R_a} \times 100 \quad (1)$$

Where $R_g$ and $R_a$ are the equilibrium resistances measured in the presence of test gas and air, respectively. The response ($\tau_{res}$) and recovery ($\tau_{rec}$) time were estimated as the time taken to raise base resistance to 90% of the maximum resistance and the time taken for resistance to fall 90% of the equilibrium resistance in the air.

2.4 Computational details

DFT calculations were realized by using Vienna ab initio simulation package (VASP)[27]. The generalized gradient approximation (GGA) with Perdew-Burke-Ernzerhof (PBE) functional was used for the exchange-correlation potential and the cut-off energy for plane-wave basis set was 500 eV [28]. The Van der Waals correction (vdW) by using DFT-D2 method was used to adsorption calculations [29]. The 4×4×1 k-point sampling was used for the first Brillouin zone (BZ) according to Monkhorst–Pack scheme[30]. In order to obtain the optimized configurations, the atoms are relaxed until the atomic force on each atom was less than 0.01 eV/ Å and the energy convergence was $10^{-6}$ eV. Bader charge analysis was used for calculating the charge transfer between absorbent surface and adsorbed molecules [31]. To simulate the ZnO surface, two ZnO layer was cleaved from its bulk wurtzite structure with the hexagonal unit cell [32] in (0001) direction with more than 20 Å vacuum space to avoid the interaction between surfaces. The O-terminated surface (000$\bar{1}$) was used to study the adsorption process and the dangling bonds on Zn terminated surface were saturated by pseudo-hydrogen atoms with Z=3/2 to prevent an unusual charge transfer to the surface[33]. After optimization of the atomic positions and the unit cell of ZnO slab, a 3×3 super cell with the lattice parameters of a=b=9.80 Å was used to perform the calculations.

## 3. Results and Discussion

3.1 Material Characterization

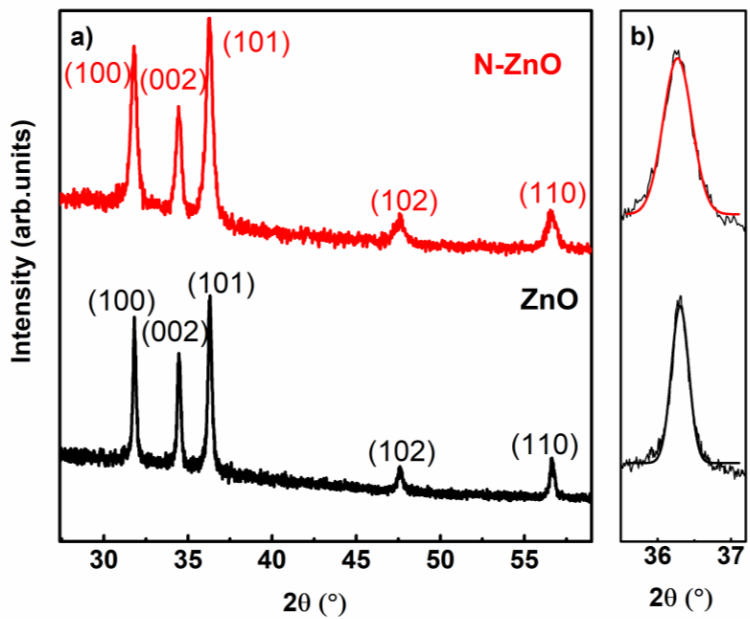

**Figure 1.** a) X-ray diffractogram of pure and nitrogen doped zinc oxide thin films and b) enlarged view of (101) peak.

X-ray diffraction (XRD) patterns of pure and nitrogen doped ZnO thin films in the ω-2θ geometry have been compared in figure 1(a), which have been indexed according to the standard JCPDS file for ZnO (36-1451). The diffraction peaks corresponding to N-ZnO indicate a hexagonal wurtzite structure similar to that of ZnO. XRD measurements confirm that incorporation of nitrogen did not give rise to the formation of any secondary phase or change in the crystal structure of ZnO. Both undoped and nitrogen doped ZnO thin films exhibit prominent diffraction peaks representing the (100), (002), and (101) planes without any distinct preferential orientation. Our results agree well with the existing literature reports of ZnO thin film grown via wet chemical route [34,35]. Such growth of ZnO thin films without any preferred orientation can be due to the random nucleation of small particles in the presence of a volatile solvent such as ethanol (boiling point~ 78 °C), which inhibits the development of grains with a definite orientation [34]. Additionally, only a minor difference in intensity of the diffraction peaks shows that the film thickness and are comparable for the ZnO and N-ZnO thin films. An enlarged view of the (101) peak for the ZnO and N-ZnO samples have been shown in figure 1(b). Clearly, (101) peak of the N-ZnO layer appears to be broader compared to the ZnO layer of comparable thickness. The full width half maximum (FWHM) of (101) peak of ZnO and N-ZnO have been estimated to be 0.20° and 0.35°, respectively. Average crystallite size of the ZnO and N-ZnO thin films have been estimated using the Scherrer equation:

$$D = \frac{K\lambda}{\beta cos(\theta)} \quad (2)$$

Where D is the crystallite size of the thin films, $\lambda$ is the X-ray wavelength (1.54 Å), $\beta$ is the FWHM in radians, K is constant with a value of 0.94, and $\theta$ is the peak position in radians. The crystallite size of the N-ZnO (~23 nm) has been found to be smaller than the ZnO film (~41 nm). This has been attributed to the addition of urea which acts as a surfactant and inhibits grain growth [23]. No notable change in peak position has been observed for the nitrogen doped film compared to the pure zinc oxide thin film, which indicates a small quantity doping of nitrogen into the lattice of ZnO [36]. Further the micro strain (ε) along the (101) plane of the thin films is calculated using the equation,

$$\varepsilon = \frac{\beta}{4\tan\theta} \quad (3)$$

Micro strain in ZnO has been found to be 2.36 x $10^{-3}$, whereas that for N-ZnO thin film is 4.56 x $10^{-3}$. Similar change in micro strain has been reported for Ag and Li doped ZnO thin films [37,38].

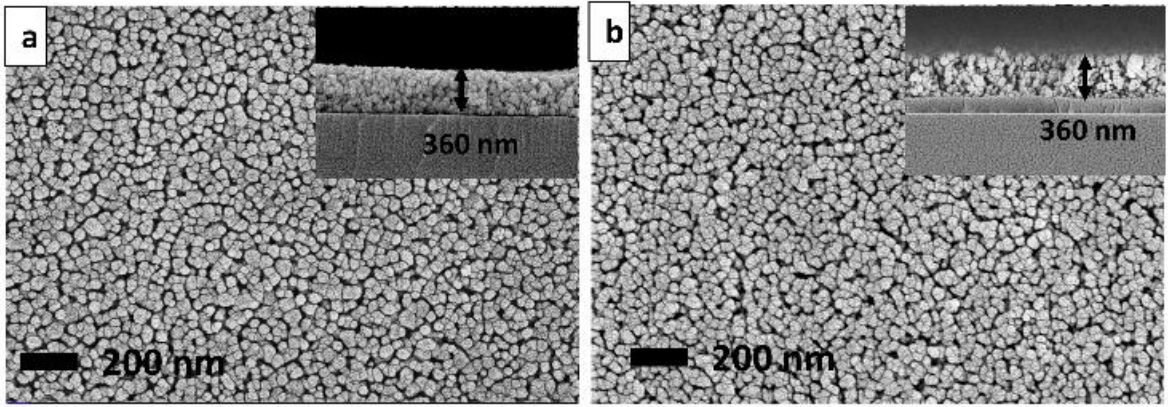

**Figure 2**. Scanning electron micrographs images of a) ZnO and b) N-ZnO thin films (corresponding cross-sectional images in inset)

Surface microstructure of the thin films have been investigated by scanning electron microscopy (SEM). Figure 2(a) and (b) represent the scanning electron micrographs of the surfaces of ZnO and N-ZnO thin films, respectively. Surface morphologies of both the samples exhibit uniform coagulated grain distribution with mesopores of average diameter around ~25 nm. The average grain size of N-ZnO (~ 30 nm) is observed to be lower than that of ZnO (~ 48 nm). From cross-sectional SEM (figure 2 inset) the thickness of the samples has been estimated to be ~360 nm.

Figure 3(a) depicts the XPS survey scan of ZnO and N-ZnO thin films. For better understanding high resolution XPS spectra has been recorded around the binding energy values corresponding to N 1s, O 1s and Zn 2p levels. As carbon 1 s peak at 284.5 eV (not shown) as reference all the binding energies have been calibrated and peak backgrounds have been corrected using Tougaard background.

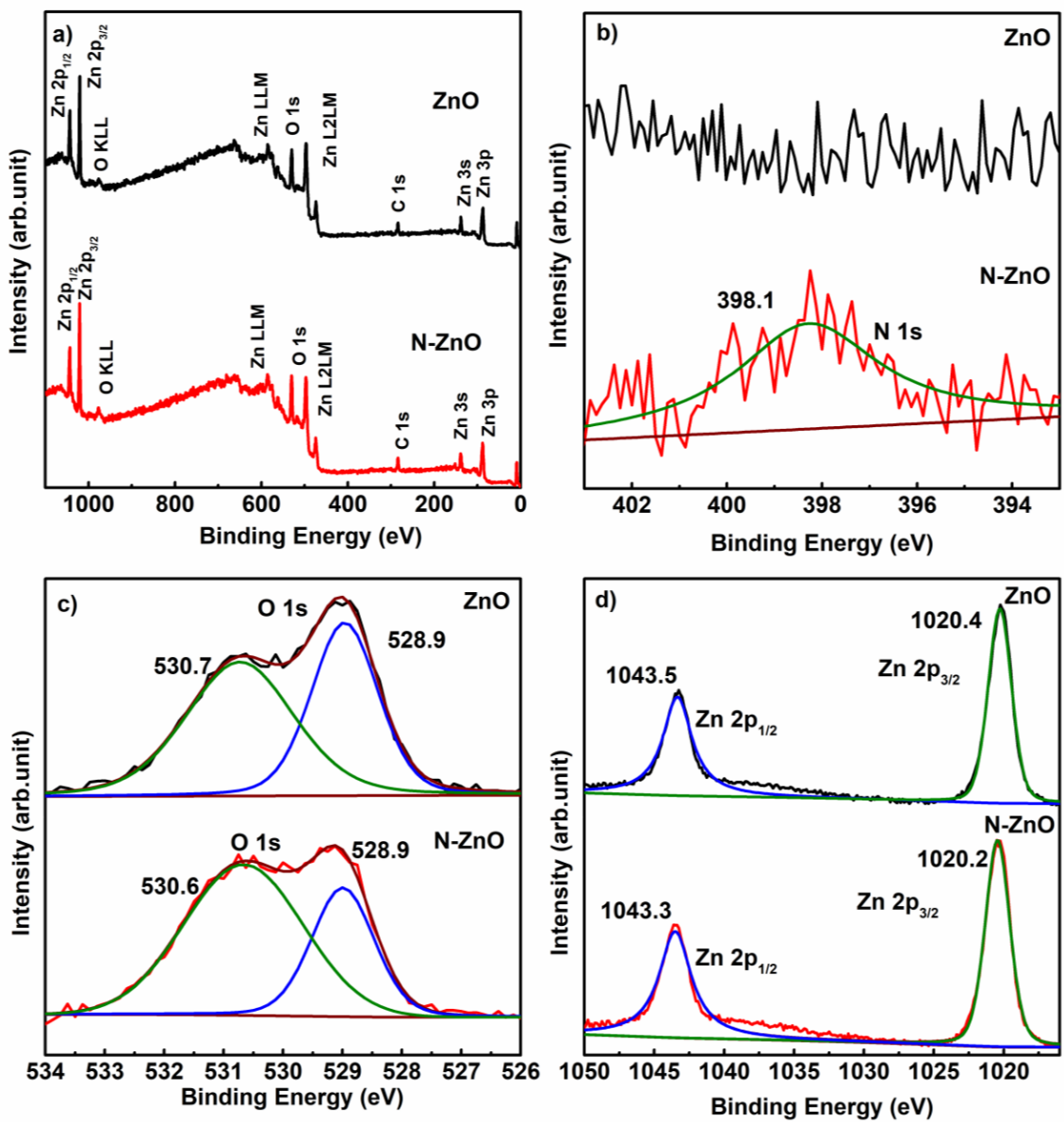

**Figure 3**. XPS a) survey scan, high-resolution scan of b) N 1s, c) O 1s and d) Zn 2p of ZnO and N-ZnO thin films.

Figure 3(b) compares the high-resolution XPS spectra of the N1s region (395-401 eV) for ZnO and N-ZnO thin films. A single peak has been observed in the N 1s region for N-ZnO, while no peak has been observed for the pristine ZnO sample. The presence of the peak confirms the effectiveness of the chosen synthesis route for the successful doping of nitrogen into ZnO. As per the theoretical predictions [39] nitrogen can be doped into ZnO by at least two different forms, e.g., N atom substitution in the oxygen site of ZnO ($N_O$), and $(N_2)_O$ is an $N_2$ molecule occupying a position on the oxygen sublattice. Usually, the substitutional N atoms exhibit binding energy of about 398 eV [23], while $(N_2)_O$ species exhibit binding energy of ~406 eV [40]. In the present study, the single peak of N1s with the binding energy of 398.1 eV confirms the manifestation of substitutional N doping in N-ZnO. This agrees well with the report of Jiao *et al.* for a small amount of N-doping in ZnO thin films [23–25]. Further, the appearance of a single peak of N1s also eliminates the possible presence of other forms of nitrogen species such as N-H, C-N, and N-N, which usually has binding energy higher than 399 eV [23,24]. Nitrogen concentration in the N-ZnO thin film has been estimated to be 0.65 at % with an N/Zn ratio of 0.045.

Figure 3(c) represents the O 1s peak corresponding to pure and nitrogen doped ZnO samples. In both cases, the O 1s peak can be deconvoluted into two peaks corresponding to Zn-O (~528.9 eV) and non-stoichiometric oxygen (~530.6 eV - 530.7 eV) [24,25]. The ratio of area under curves $A_V/A_L$ ($A_V$ and $A_L$ are the area under the curve for non-stoichiometric and lattice oxygen respectively) for ZnO and N-ZnO is 0.92 and 1.61 respectively, which clearly indicates that oxygen vacancy or oxygen non-stoichiometry in ZnO increases with N doping. This observation agrees well with previous literature reports[24,25]. The substitutional N (-3) doping in O (-2) lattice creates a charge imbalance, which is compensated by forming oxygen vacancy (+2) [41]. Figure 3 (d) displays the Zn 3p peak for ZnO and N-ZnO thin films, which exhibit the doublet peaks of Zn corresponding to 3p$_{3/2}$ and 3p$_{1/2}$. The Zn 3p$_{3/2}$ and 3p$_{1/2}$ peaks of ZnO appear at 1020.4 eV and 1043.5 eV, respectively, while in case of the N-ZnO thin film the Zn 3p peaks exhibit a slight shift of 0.2 eV towards lower binding energies (1020.2 eV and 1043.3 eV, respectively). Considering the fact that the O1s core level corresponding to the lattice oxygen is the same for both the samples, shift of Zn 3P towards lower binding energy indicates that the electron density of Zn in the N-ZnO sample is slightly higher, which can be attributed to the formation of Zn-N bonds [25].

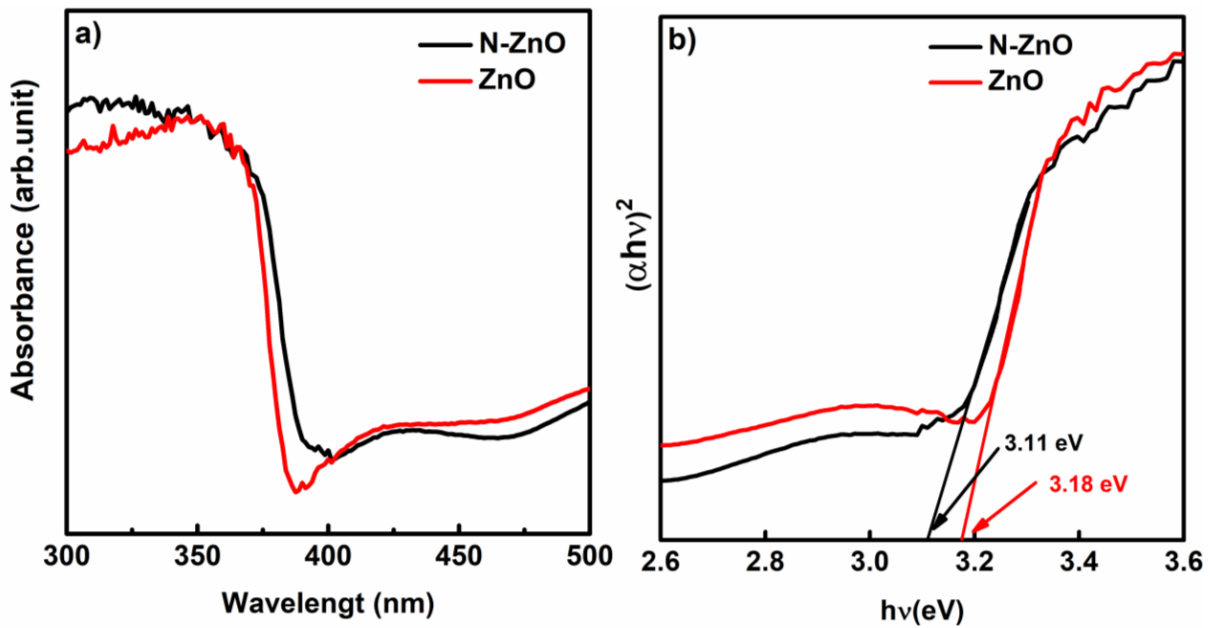

**Figure 4**. a) Diffuse reflectance spectroscopy and b) corresponding Tauc's plot of ZnO and N-ZnO thin films.

UV-Visible diffuse reflectance technique has been utilized to evaluate the effect of nitrogen incorporation on ZnO's bandgap, and it is depicted in figure 4. The DRS spectrum (Figure 4.a) has shown a slight red shift in absorbance with N doping in ZnO. The bandgap estimated from the Tauc plot (Figure 4.b) is 3.12 eV for N-ZnO thin film, lower than that of ZnO thin film (3.18 eV).

The reduction in band gap is attributed to the contribution of high lying N-2p orbitals in addition to the O-2p orbitals in forming the valence band edge of the N-ZnO film [42,43].

3.2 Gas sensing characteristics

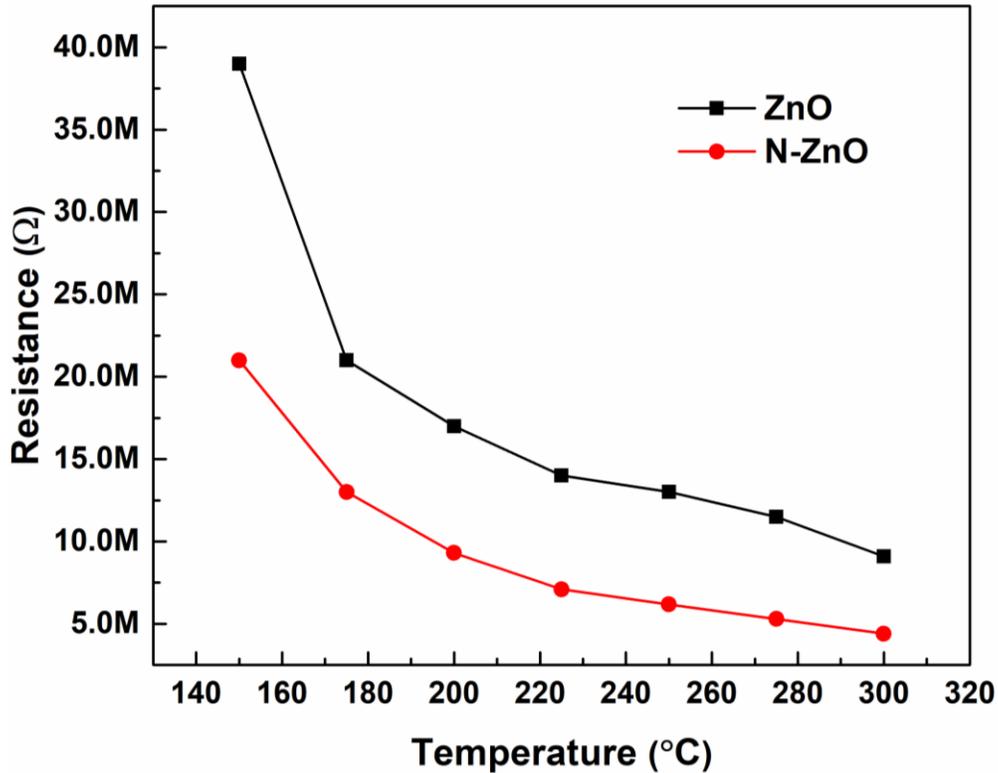

**Figure 5.** Base resistance of thin films at various temperatures in ambient conditions

The base resistance of thin films has been measured at various temperatures and the same is illustrated in figure 5. The thin films have shown the typical characteristics of a semiconducting material, where the resistance decreases with increase in temperature. The N-ZnO thin film were found to be less resistive compared to ZnO thin films. The decrease in resistance of nitrogen doped film is attributed to increased oxygen vacancy ($V_o^{2+}$) [41,44], which is a shallow donor in ZnO. The excess availability of $V_o^{2+}$ increases the carrier concentration and decreases the resistance of N-ZnO thin film at all the working temperatures.

Ethanol sensing characteristics of the N-ZnO and pristine ZnO thin films have been compared in the temperature range of 100 °C- 300 °C. Figure 6(a) depicts the temperature-dependent sensing responses (%) of N-ZnO and ZnO thin films of the comparable thickness (~ 360 nm) towards ethanol concentration of 300 ppm. The sensing response of the N-ZnO thin film is much higher than that of the pristine ZnO thin film sensor over the entire sensor operating temperature range. The ZnO thin film sensor exhibits the highest response of ~81.2% at an optimum operating temperature of 250 °C above which the sensor response decreases. In contrast, the N-ZnO film exhibits a response ~99.7% at a lower optimum operating temperature of 225 °C. Unlike the ZnO film, the N-ZnO film's sensing response remains relatively constant (>99%) for the operating temperature ranging between 225-300 °C. Further, the N-ZnO thin-film sensor exhibits a significant response of 45.5 % at an operating temperature of as low as 100 °C at which the ZnO thin film does not display any detectable sensing response towards ethanol vapor.

Figure 6(b) displays a representative graph depicting resistance transient for both the sensors at an ethanol concentration of 300 ppm and a working temperature of 225 °C. The N-ZnO film has shown significantly faster response time (~12 s) time compared to ZnO film (~33 s), while both films have shown comparable recovery times of ~360 seconds. The films' response time is observed to be faster than recovery time, which is a typical characteristic of n-type metal oxides [45,46]. The same trend is observed for static gas sensing setups where the recovery is due to the lower airflow over the films [26]. The influence of atmospheric humidity in the ethanol sensing characteristics has been carried out over a relative humidity (RH) range of 20 % - 80%.

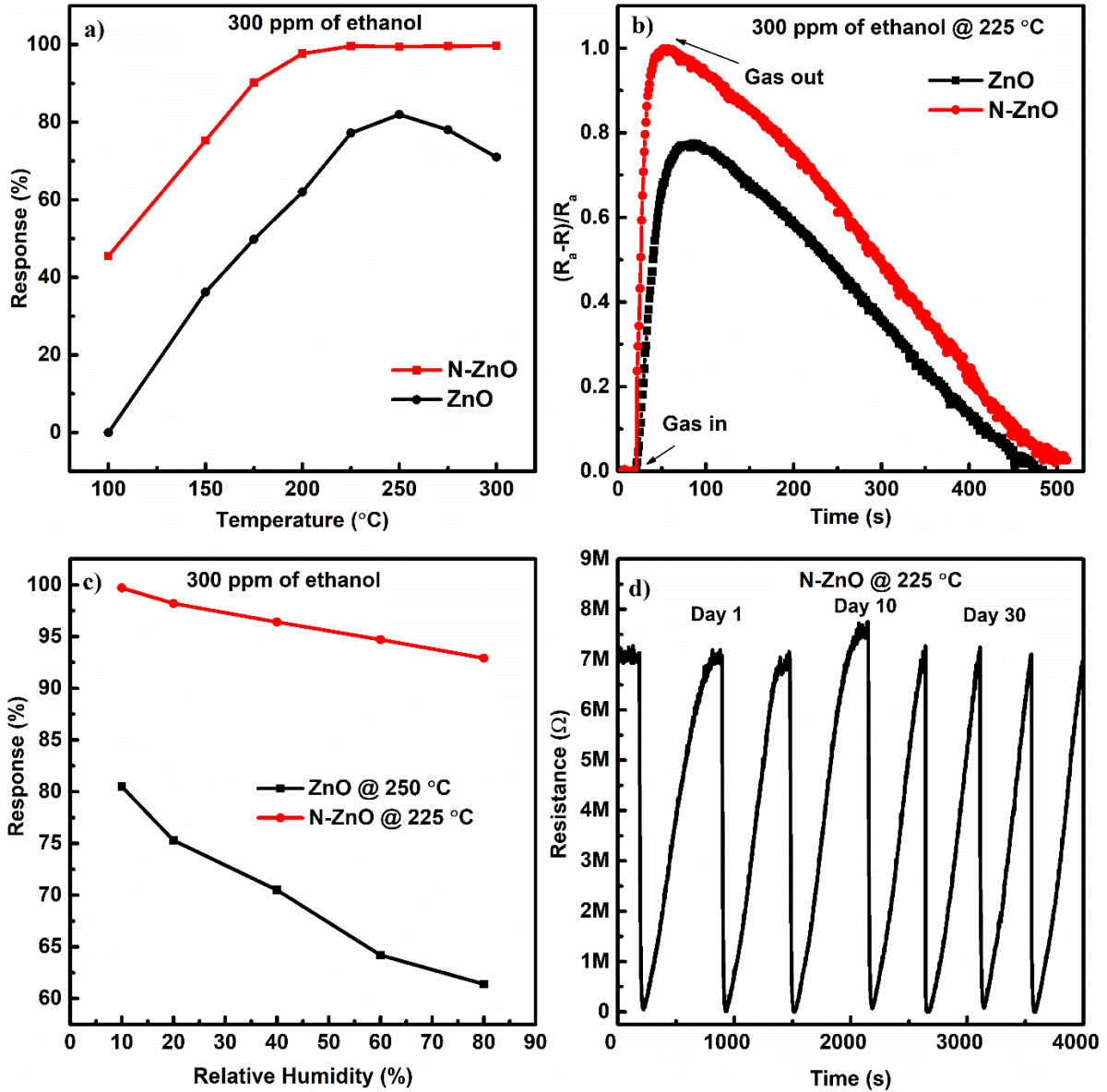

**Figure 6**.a) Ethanol sensing characteristics at a concentration of 300 ppm at different temperatures b) sensor response transient at the optimum temperature of 225 °C, c) RH dependent response for the ZnO and N-ZnO sensors and d) extended stability resistance transient of N-ZnO sensor at 225 °C. Lines are to guide the eyes.

Figure 6(c) compares the RH dependence of ethanol sensing response of the N-ZnO and ZnO thin films. The sensor response at different RH has been set and measured at the optimum working temperature of 250 °C and 225 °C for ZnO and N-ZnO thin films respectively. A ~ 25 % drop in response has been observed in the entire range of RH for ZnO. On the other hand, N-ZnO demonstrates fairly steady response (reduction of only 6 %) in the given range of RH. This indicates that nitrogen doping in ZnO thin film inhibits adsorption of ambient moisture on the film's surface which aids in superior ethanol sensing response. This observation agrees well with the theoretical investigation of Tit *et al.* who demonstrated that water adsorption on ZnO surface has reduces on incorporation of nitrogen into the ZnO lattice [47,48]. Figure 6(d) represents response of nitrogen doped zinc oxide thin film for 300 ppm of ethanol at 225 °C over the period of one month. The sensor exhibits stable response with good base line recovery during this time.

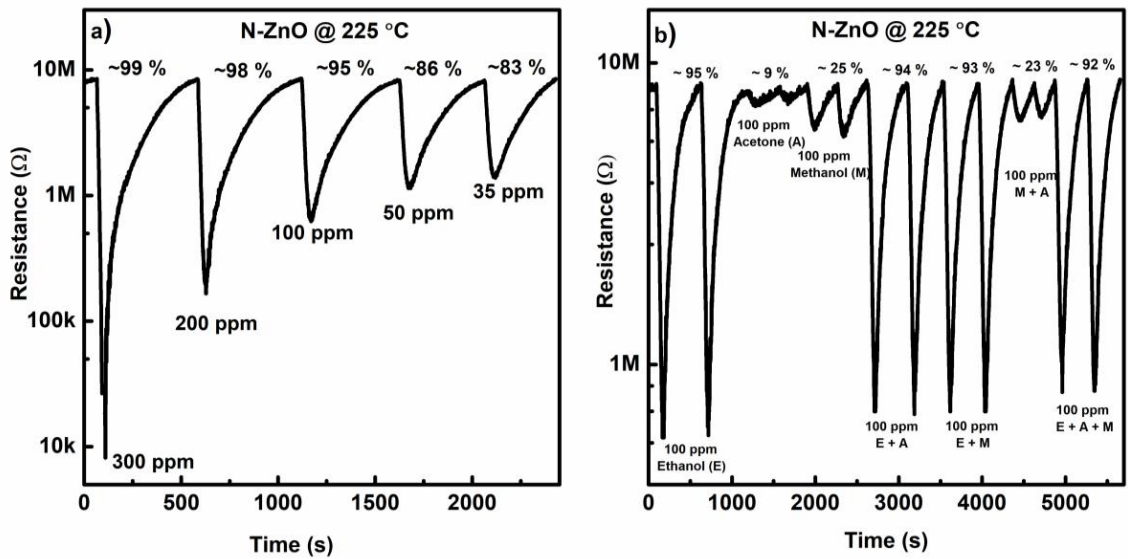

**Figure 7**. a) Concentration variation profile of N-ZnO thin film at 225 °C b) sensing response towards 100 ppm ethanol, acetone, methanol and their mixed condition for N-ZnO thin film at 225 °C. (E-ethanol, M-Methanol and A-Acetone).

Ethanol sensing response of the N-ZnO thin film have been further investigated over a wide range of concentration. Figure 7(a) portrays the resistance transient of the N-ZnO thin film sensor over ethanol concentration range of 35-300 ppm at its optimum temperature (225 °C). At the lowest measured concentration of 35 ppm, the N-ZnO sensor has exhibited a high response of ∼83.7%, while the pure ZnO sensor has an inferior performance (∼53 % at 35 ppm). The sensor response, S, is given by

$$S = \beta C^n \quad (5)$$

Where C is the concentration, n is the sensitivity, and β is the sensitivity coefficient. The sensitivity coefficient is a function of operating temperature and transduction activation energy [45,49]. The sensitivity, n, for ZnO and N- ZnO sensors has been estimated to be ~0.19 and ~0.08, respectively (Figure S1), indicating a higher response of N-ZnO results from the facile transduction due to improved receptor function of the sensor [50]. Further, sensing response of the ZnO and N-ZnO thin films towards ethanol have been compared to their sensing response towards methanol and acetone vapour. Methanol is a common additive in ethanol to denature it for laboratory and industrial applications. Figure 7(b) shows selectivity of ethanol against methanol, acetone and selective sensing of ethanol in mixed conditions. The sensing response exhibits that N-ZnO offers superior selectivity towards ethanol (~95 %) in comparison with methanol (~25 %) and acetone (~9 %). The higher selectivity of nitrogen doped ZnO sensor towards ethanol can be attributed to the higher acidity of ethanol compared to other two gases. The acidity (Brønsted acid) of ethanol, methanol and acetone are 15.5, 15.7 and 19.3 respectively [51]. Hence ethanol (stronger Brønsted acid) can consume more adsorbed oxygen from a basic oxide surface, since nitrogen doping in general leads to pronounced basic character in metal oxides [52] . This leads higher sensing response of ethanol compared to other two gases. The N-ZnO sensor also demonstrated effective sensing of ethanol in the mixed ethanol-methanol condition with a marginal reduction in sensor response. This indicates the capacity of N-ZnO to sense ethanol in the mixed gas environment.

Additionally, the N-ZnO thin film have displayed stable sensing towards ethanol under the interference of methanol and acetone.

Our experimental results indicate that the enhanced ethanol sensing performance of N-ZnO originates from the increased concentration of oxygen vacancy and modified receptor function. In order to gain further understanding of improved sensitivity and selectivity of N-ZnO thin film, DFT study has been conducted and it is described in the following section.

3.3 DFT study

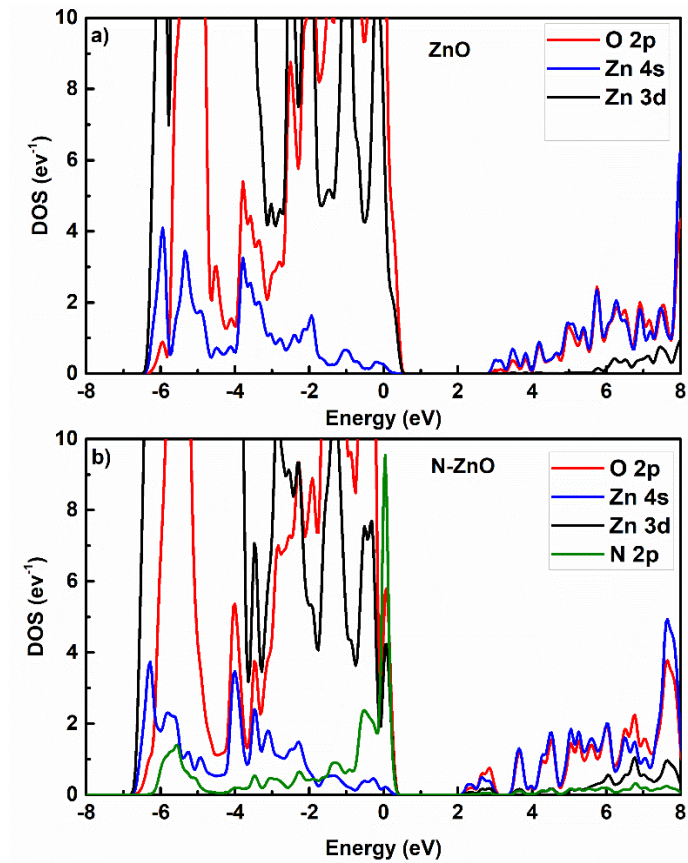

**Figure 8.** Density of states of a) ZnO and b) N-ZnO slabs

The optimized structures of ZnO and N-ZnO used for DFT study has been portrayed and described in the supplementary information (figure S2). To study the electronic structure, the density of states (DOS) for pristine ZnO and N doped ZnO with O vacancy slabs have been calculated. As shown in figure 8(a), the DOS of ZnO around the Fermi level consist of O 2p states and Zn 3d states and there is an energy gap of 2.1 eV above the Fermi level, which is lower that experimental band gap reported in literature [25]. This is a common shortcoming in DFT calculation were in which the lowest unoccupied orbital is underestimated [53]. The conduction band edge is made up of Zn 4s and O 2p, which agrees well with previously reported literature [54,55]. Figure 8(b) illustrates the PDOS of N-ZnO, which have a pick on Fermi level arises from the N-p states and the energy gap above Fermi level has been dropped to 1.4 eV. The reason for reduction in band gap being the electronic states appeared between 2.2 eV and 3.2 eV due to oxygen vacancy.

To simulate the adsorption behavior of ethanol molecule on ZnO and N-ZnO, adsorption of ethanol on O-site of ZnO surface and O-site and N-site of N-ZnO surface have been considered. Figure 9 (a, b &c) shows the optimized structures of adsorbed molecules on the surfaces. In both cases of adsorption on O-site, the H atom of the hydroxyl group of ethanol (H(m)), is bonded to the O atom of the oxide surface (O(s)) by hydrogen bonding, which is in agreement with previously reported literature[56]. On the other hand, the ethanol adsorption on the N-site of the N-ZnO surface is by interaction of the O atom of ethanol's hydroxyl group (O(m)) and the surface's N atom (N(s)). The adsorption/binding energy, ΔE, is calculated by the following equation:

$$\Delta E = [E_{molecule+surface} - (E_{molecule} + E_{surface})] \qquad (6)$$

where $E_{molecule + surface}$, $E_{molecule}$ and $E_{surface}$ are the total energy of the adsorbed molecule + surface complex (here ethanol + ZnO or N-ZnO slabs), the total energy of the ethanol molecule in the same cell and the total energy of the same clean ZnO and N-ZnO surfaces, respectively. The estimated bond length and adsorption energies are presented in Table 1.

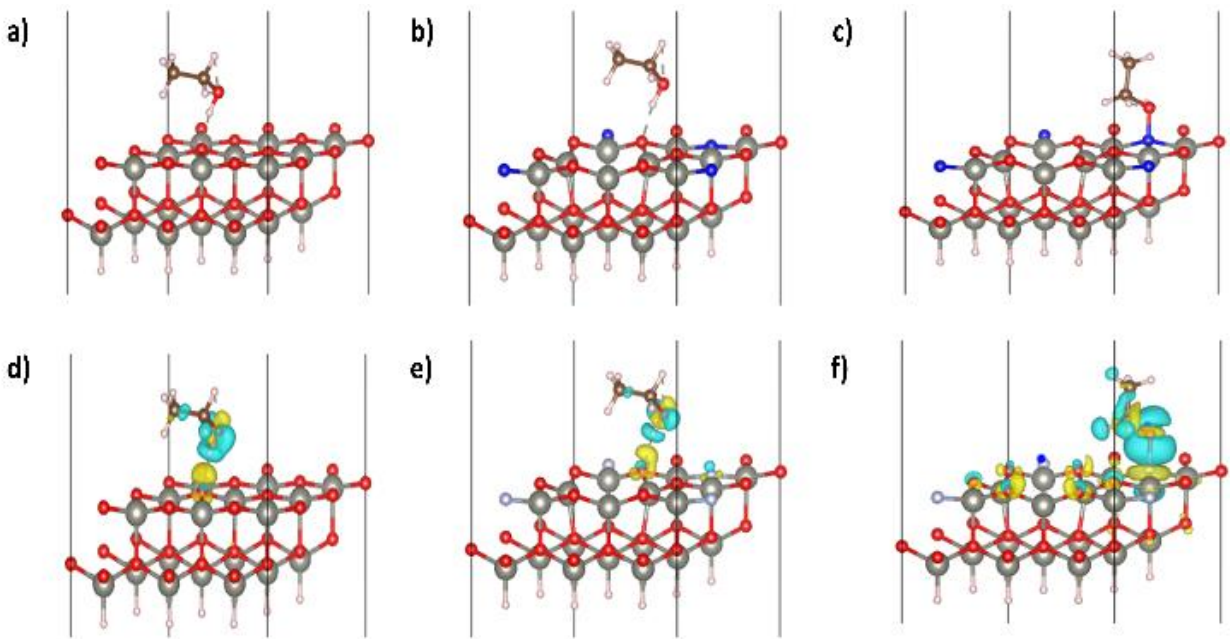

**Figure 9**. Optimized geometry of ethanol adsorbed on a) O-site of ZnO surface b) O-site of N-ZnO surface c) N-site of N-ZnO surfaces and charge density differences plots of ethanol adsorbed on d) O-site of ZnO surface e) O-site of N-ZnO surface f) N-site of N-ZnO surfaces. The yellow color shows electron accumulation and blue color shows electron depletion. (O atoms are red, Zn atoms are ash, N atoms are blue, H atom are white, C atom are brown)

It has been observed that the adsorption processes are in accordance with energy release because of negative adsorption energies. Therefore, the adsorption of ethanol on the considered surfaces is exothermic and the final complexes from the energetic consideration (or enthalpic effects) are stable. The results also show that the energy variations for the adsorption of ethanol on the O-site of considered surfaces are −0.64 and −0.60 eV for ZnO and N-ZnO surfaces, respectively but for N-site of N-ZnO it is -1.65 eV. It may be concluded that the adsorption of ethanol on N-sit of N-ZnO is chemisorption, however electronic density or charge difference analysis would be used for the detailed description of the nature of adsorption in the upcoming sections.

**Table 1**. Adsorption energy, bond distance and charge transfer of ethanol and methanol at various sites.

| Adsorption site | Adsorption energy (eV) | Bond distance (Å) | Charge transfer, Δq (e) |
|---|---|---|---|
| Ethanol on O-site of ZnO | -0.64 | 1.64 | 0.20 |
| Ethanol on O-site of N-ZnO | -0.60 | 1.70 | 0.13 |
| Ethanol on N-site of N-ZnO | -1.65 | 1.56 | 0.46 |
| Methanol on O-site of N-ZnO | -0.42 | 1.73 | 0.11 |
| Methanol on N-site of N-ZnO | -1.45 | 1.57 | 0.43 |

The charge density differences (CDD) for the adsorption of ethanol on the considered surfaces has been estimated by:

$$\Delta \rho = \rho_{molecule+surface} - (\rho_{molecule} + \rho_{surface}) \quad (7)$$

Where, Δρ is the charge density difference, $\rho_{molecule+surface}$, $\rho_{molecule}$ and $\rho_{surface}$ are the charge densities of the complex of ethanol + surface, the ethanol molecule and clean surface, respectively. As it is shown in Figure 9(d & e), the O-site and the region around it in both surfaces has a positive charge density difference (CDD) due to accumulation of electrons (yellow iso-surfaces) and the region around the H atom of ethanol is experienced a charge density depletion confirming the hydrogen bonding. Also Figure 9(f) represents that the N-site of the N-ZnO surface gains charge density and there is a strong charge density depletion around the O atom of ethanol which shows the strong interaction between the N and O atoms. The Bader charge analysis also shows that in all cases, charge transfers from ethanol to the surface and therefore the surface may be considered as an acceptor. The calculated charge transfers from ethanol to the surface, Δq, are collected as Table 1. The Δq is calculated by considering the charges on individual atoms on the surface after ethanol adsorption and which has been described in the supplementary information (Figure S3 and Table S1). These results indicate the charge transfer magnitude in the adsorption in N-site is significantly larger than other two cases. The PDOS for the atoms before and after adsorption of ethanol is portrayed in figure S4 (a -f). The results indicate a strong change in PDOS with ethanol adsorption in N-site of N-ZnO compared to O site adsorption of ZnO and N-ZnO. This shows that N(s) may have new hybridization with the O(m) states in these energies.

3.4 Ethanol sensing mechanism

The sensitivity of a sensor depends on adsorption-desorption and interaction of target gas with adsorbed oxanions[15,21]. The nature of oxanions depends on the sensor working temperature, that is, $O^{2-}$(above 300 °C), $O^-$ (above 200 °C and below 300 °C ) and $O_2^-$ (below 200 °C) [21]. In the present investigation at the sensing temperature of 225 °C, the most prominent absorbed oxygen species on the sensor surface is of $O^-$. The proposed mechanism for ethanol sensing is shown figure 10. When the N-ZnO sensor is heated in the atmospheric conditions, gaseous oxygen molecules get adsorbed as oxanions as shown in step 1. The oxanions accepts electrons from the surface to create a depletion layer and the conductivity of the sensor decreases. Ethanol gets adsorbed to the sensor surface once it is introduced.

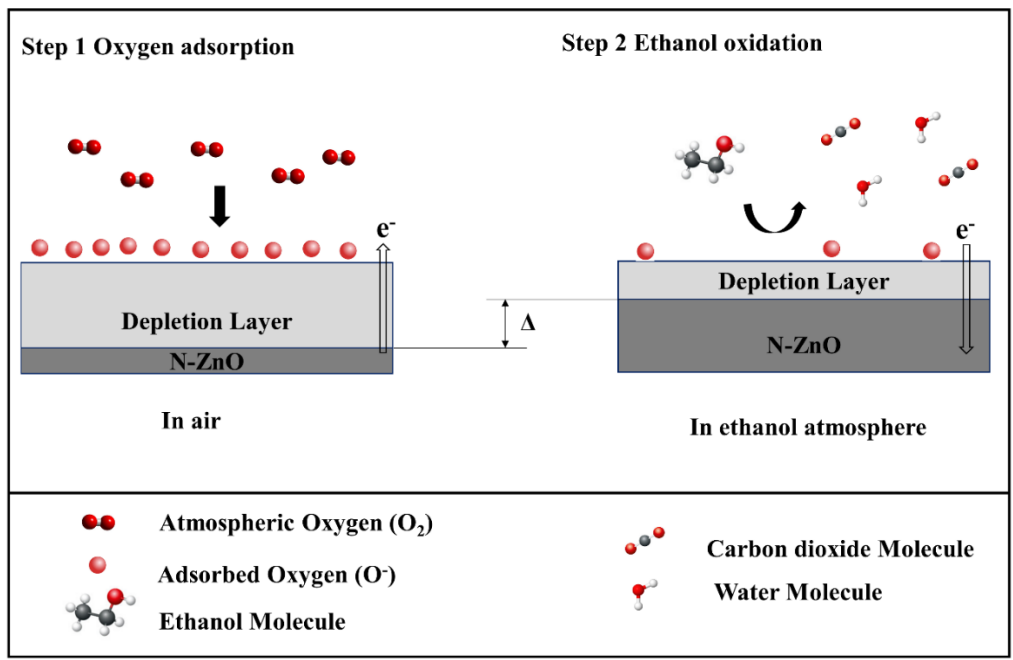

Figure 10. Proposed ethanol sensing mechanism of N-ZnO thin film

The adsorbed ethanol molecule undergoes oxidation (step 2) to form water and carbon dioxide following the reaction,

$$C_2H_5OH + 5O^- \rightarrow 2CO_2 + 2H_2O + 5e^- \qquad (8)$$

which results in release of electrons into the conduction band of N-ZnO and increases its surface conductivity. The width of depletion layer depends on the amount surface absorbed oxanions. The change in the depletion width as shown in step 2 (figure 10) is quantified as change in resistance, which is used to estimate the sensor response. Stronger adsorption energy of ethanol on N site of N-ZnO surface facilitates easy adsorption and further decomposition of ethanol. Moreover, nitrogen doping increases oxygen vacancy in ZnO (from XPS) and oxygen vacancies are reported to favour dissociative adsorption of atmospheric oxygen and thus improving the gas sensing abilities of metal oxides [11]. Further preferential adsorption of oxygen molecules compared to water molecules (RH variation study) also contributes to enhanced sensitivity of N-ZnO compare to ZnO [26,47]. Lastly, the smaller grains of N-ZnO thin films than ZnO thin film ( shown in figure 2) have also added to the enhanced ethanol sensing of N-ZnO [5,16]

We have compared our results with reported literature and it is shown Table S2.

## 4. Conclusion

To summarise, ZnO and N-ZnO thin films have been synthesized by sol-gel technique using spin coating on $SiO_2$/Si substrates. Both the films have been found to be crystalline and they have the same wurtzite structure. Additionally, the surface of the films was observed to be mesopores. Substitutional doping of nitrogen was confirmed via x-ray photoelectron spectroscopy analysis. Nitrogen doping in ZnO led to increase in the concentration of oxygen vacancies, that facilitates adsorption of oxanions on the sensor surface. Ethanol sensing characteristics of the samples have been studied by varying the operating temperature (100 °C – 300 °C) and concentration of the ethanol vapour (35- 300 ppm). N-ZnO film exhibited enhanced sensing response (∼99.7 %), faster response time (12 S) and a lower operating temperature (225 °C) compared to the ZnO film. Further, the N-ZnO thin film exhibited reasonable sensing response (~45.5 %) towards ethanol vapour (300ppm) at an operating temperature as low as 100 °C. The N-ZnO sample exhibited superior resistance towards ambient moisture, long term stability and selectivity towards ethanol compared to methanol. In order to elucidate the microscopic mechanism of ethanol sensing characteristics of N-ZnO and ZnO, charge transfer between absorbent surface and adsorbed molecules have been estimated using first principal DFT calculations. Analysis of the PDOS and Bader charge indicated facile adsorption of the ethanol molecules on the N-sites of N-ZnO. Based on these results the improved ethanol sensing characteristics of N-ZnO thin film have been attributed to its modified receptor function due to nitrogen doping.

## 5. Acknowledgment

PKS and ARC acknowledge the Ministry of Human Resource Development, India (SPARC) for partial financial support of the work (vide Letter No. SPARC/2018-2019/P252/SL dated 15-03-2019), the Central Research Facility (CRF) of Indian Institute of Technology Kharagpur for various characterization supports for the synthesized material, SEM facility under DST-FIST program at Materials Science Centre and Prof. Subhasish Basu Majumder, Materials Science Centre, for extending experimental facilities pertinent to gas sensing measurements.